\documentstyle[12pt]{article}

\def\mo{micro-organism}
\def\cOa{+{\cal O}(\alpha)}
\def\v{{\bf v}}

\def\nn{\nonumber \\}
\def\e{{\rm e}}
\def\ss{shape space}
\def\po{{\cal P}}
\def\ef{efficiency}

\begin{document}
\begin{flushright} 
OCHA-PP-71 \\
January 1996
\end{flushright}

\vfill

\begin{center}
{\LARGE\bf 
Efficiency of swimming of \mo\ and singularity in shape space}

\vfill

{\normalsize Masako KAWAMURA\footnote{E-mail: masako@phys.ocha.ac.jp} }
\vfill
{\it
Department of Physics, Faculty of Science, 
Ochanomizu University

1-1, Otsuka 2, Bunkyou-ku, Tokyo, 112, JAPAN}

\end{center}

\vfill

\begin{abstract}
Micro-organisms can be classified into three different 
types according to their size.
We study the \ef\ of the swimming of \mo\ in two dimensional fluid 
as a device for helping the 
explanation of this hierarchy in the size.
We show that the \ef\ of flagellate becomes unboundedly large, 
whereas that of ciliate has the upper bound.
The unboundedness is related to the curious feature of the \ss, that is,
 a singularity at the basic shape of flagellate.
\end{abstract}
\vfill
\newpage

%%%%%%%%%%%%%%%%%%%%%%%%%%%%%%%%%%%%%
\section{Introduction}

One thing that \mo\ attracts us is its specificity:
The environment in which they live is so sticky that an ordinary 
Newtonian mechanics does not apply to their motions.
Surprisingly, however, the exploration of the mathematical structure 
which lies behind the swimming of \mo\ has led to 
the gauge theory and the string theory.
Shapere and Wilczek discussed the problem of the 
swimming of \mo \ in the light of the gauge field theory\cite{wil1}.
They took up the motion of ciliate in two and three dimensional fluid.
Inspired by their works, not only the velocity of 
ciliate, but also that of flagellate in two dimensional fluid were obtained 
and their motions were studied from the point of view of string in 
Ref.\cite{ma}.
It is known that \mo s can be grouped into three different types 
according to their size or the 
way of swimming, {\it i.e.}, ciliate, flagellate and bacterium 
with bacterial flagella\cite{2}.
In particular, such a kind of bacterium can exist only in three 
dimensional fluid, whereas 
ciliate and flagellate can exist even in two dimensional fluid.
The difference in their {\it shape} in two dimensional fluid was discussed 
with the help of representation of $W_{1+\infty}$ algebra in Ref.\cite{ma2}.
We now turn our attention to the {\it size} of \mo.
The length scale of ciliate, such as paramecium, is 
$20 \sim 2 \times 10^4 \mu {\rm m}$.
The size of flagellate, such as sperm, 
is smaller than ciliate, $1 \sim 50 \mu {\rm m}$.
Bacterium with bacterial flagella is even smaller, $0.2 \sim 5 \mu{\rm m}$.
Thus, we can find a hierarchy of the size of \mo\ and 
this is the prime focus in this paper.

Here, we shall discuss the problems in the efficiency of swimming of \mo 
\ in two dimensional fluid.
In particular, we would like to investigate whether the calculation 
of the efficiency is useful as a tool for facilitating 
the explanation of the problem on the
hierarchy of the size.
In order to discuss the swimming of \mo, we need to know fluid mechanics 
at low Reynolds number.
In the next section, we will give a brief review of the fluid mechanics at 
low Reynolds number and some useful results.
We also need to know the power expenditure of \mo \ to define 
the efficiency of the swimming of \mo.
In section 3, following \cite{wil2}, 
we shall show that we can define the metric tensor on shape space from 
the power expenditure.
In section 4, we will formulate the efficiency of the swimming of \mo \ 
in the limit of Reynolds number zero and find that the \ef\ of flagellate 
becomes unboundedly large, whereas that of ciliate has the upper bound.
The result we obtain sheds some light on the question of the 
hierarchy of the size of \mo.
It is thus natural to ask whether there is a relationship between 
the structure of the \ss \ and the problem of the efficiency.
This is briefly explored in section 5, where it is shown that the 
considerations of the metric tensor in \ss \ lead to an appealing result, 
that is, a singularity arises at the basic shape of flagellate.
The singularity makes it possible to decrease the power expenditure 
and to increase the \ef\ infinitely.
The last section is devoted to summary.

%%%%%%%%%%%%%%%%%%%%%%%%%%%%%%%%%%%%%%%%%%%%%%%%%%%%%%%%%%%%%%%%
\section{Fluid Mechanics at low Reynolds number}

In order to discuss the motion of \mo, we need to consider 
the fluid mechanics of low Reynolds number.
Reynolds number $Re$ is defined as the ratio of the inertial force to the 
viscous force,
\begin{equation}
Re = \frac{\rho V L}{\mu}, \label{re}
\end{equation}
where $\rho$, $V$, $L$ and $\mu$ denote the density of the fluid, 
the typical velocity, the typical length scale and the coefficient 
of viscosity of the fluid, respectively.
Since for given $\rho$ and $\mu$, the size and the velocity of \mo\ are 
exceedingly small, Reynolds number becomes small.
Requirement of imcompressible flow and low Reynolds number lead to the
following equations of motion for the fluid velocity $\v$ and the pressure $p$:
\begin{equation}
\nabla\cdot \v=0 
\end{equation}
\begin{equation}
-{1 \over \mu}\nabla p +\Delta \v =0
\end{equation}
Notice that these equations are linear and do not include time-dependent terms.
Therefore, if the swimmer changes its shape periodically,
the net motion due to one cycle of the deformation is invariant 
under the arbitrary time rescaling.
It only depends on the geometry of the sequence of the deformation.
Therefore, the time $t$ is just a parameter that labels the sequence of 
the deformation.
A \mo \ appears through a boundary condition.
If we denote the surface of \mo s as 
\[
S = S(\sigma,t),
\]
where $\sigma_i{\rm 's}\ (i=1,2,\cdots,D-1)$ parametrize the surface 
of $D$-dimensional \mo,
then the boundary condition becomes 
\begin{equation}
\v (S)= \frac{\partial S(\sigma, t)}{\partial t}.
\end{equation}
This is the no-slip boundary condition.
We assume that $S$ is a periodic function of $t$ with period $T$,
\[
S(\sigma,t)=S(\sigma, t+T),
\]
and the swimmer changes its shape infinitesimally from the basic shape $S_0$,
\begin{eqnarray}
S(t) & = & S_0 + \alpha(t), \hspace{0.7cm}  
|\alpha(t)| \ll |S_0| \\
\alpha(t) & = & \sum_n \alpha_n w_n,
\end{eqnarray}
where $w_n$ is a complete set of the functions describing the 
deformation of the shape.
In computing a velocity of the swimmer, we can exploit the 
infinitesimalness of  $\alpha$ by using perturbation theory.
Then we expand the net motion due to one cycle of the deformation 
up to the second order of $\alpha$ and obtain, 
\begin{equation}
\frac{1}{2}\int_0^T \sum_{n,m}F_{mn} \alpha_m \dot{\alpha}_n dt,
\end{equation}
where 
\[ F_{mn} = T^i F^{T^i}_{mn} + M^{ij}F^{M^{ij}}_{mn},
 \]
and $T^i, \ M^{ij}$ are translation and rotation operator 
acting on the shape space, respectively.
The first-order term does not contribute to the net motion because the term is 
a total derivative.
$F_{mn}$ is interpreted as field strength tensor, evaluated at a 
shape $S_0$
(See \cite{wil1}).

%%%%%%%%%%%%%%%%%%%%%%%%%%%%%%%%%%%%%%%%%%%%%%%%%%%%%%%%%%%%%%%%%%
\section{Metric tensor on shape space}

A shape space is a set of all possible shapes $S_0$.
It is convenient to consider the metric tensor on shape space, 
which was proposed by Shapere and Wilczek \cite{wil2}, 
before moving on to the problem of the efficiency of swimming of 
\mo s.
The power expenditure required to deform the body of \mo \ is given by 
\begin{equation}
\po = - \int_{S_0} v_i \sigma_{ij} dS_j,
\end{equation}
where $\sigma_{ij}$ is the stress tensor,
\begin{equation}
\sigma_{ij} = -p \delta_{ij} + \mu \left( 
\frac{\partial v_i}{\partial x_j} + \frac{\partial v_j}{\partial x_i} 
\right),
\end{equation}
and we performed the integration over the surface of \mo \ $S_0$.
It can be represented as 
\[
\po = 4\pi\mu \sum_{m,n}P_{mn}\dot{\alpha}_m \dot{\alpha}_n
\]
Note that this formulation is only applicable to the case of Reynolds 
number zero.
Moreover, $\po$ is non-negative quantity, since we can rewrite $\po$ as 
\[
\frac{1}{2}\mu \int_{{\rm exterior\  of\ }S_0}
(\partial_j v_i + \partial_i v_j)^2 \, dV      
\]
by using Gauss' theorem and the equations of motion.
$\po$ is also invariant under the 
transformation of reference axes for $\alpha \rightarrow 
\tilde{\alpha}(\alpha)$. 
Thus we can express $\po$ in terms of the invariant line element $ds$
in \ss:
\[
\left({ds \over dt}\right)^2=\po
\]
Therefore, $P_{mn}$ can be interpreted as the metric tensor evaluated at 
$S_0$.

%%%%%%%%%%%%%%%%%%%%%%%%%%%%%%%%%%%%%%%%%%%%%%%%%%%%%%%%%%%%%%%%%%
\section{The efficiency of swimming of \mo}

Let us define the efficiency of the swimming of 
\mo\ as the ratio of the absolute 
value of the average velocity to the average power expenditure.
\begin{equation}
\eta \equiv \frac{|\bar{U}|}{\bar{\cal P}}, \label{defeff}
\end{equation}
where 
\begin{equation}
 \bar{U}  =  \frac{1}{T} \int_0^T U(t) dt
= \frac{1}{2T} \int_0^T \sum_{mn}F_{mn} \alpha_m \dot{\alpha}_n dt, 
\end{equation}
\begin{equation}
 \bar{\cal P}  =  \frac{1}{T} \int_0^T {\cal P} dt 
= \frac{1}{T} \int_0^T \sum_{mn}P_{mn} \dot{\alpha}_m 
\dot{\alpha}_n dt, 
\end{equation}
and $U(t)$ is the velocity of \mo \ at time $t$.
We shall assume that the motion of \mo\ does not include rotational 
one which decreases the \ef, since the shortest distance between two points 
is a straight line and our interest is in the maximum value of the \ef.

We now employ variational method to calculate the maximum value of the 
efficiency $\eta$:
\begin{equation}
\frac{\delta \eta}{\delta \alpha_m} = 0 \label{vari}
\end{equation}
Henceforth, we shall concentrate on \mo \ in two dimensional fluid,
{\it i.e.}, ciliate and flagellate, and compare the \ef\ of ciliate 
with that of flagellate.

%%%%%%%
\subsection{Ciliated motion}

We can take advantage of the two-dimensionality by using 
the complex coordinate.
We regard the shape of ciliate as infinitesimal deformation of a circle:
\[
S(t,\theta)=\e^{i\theta}+\sum_n\alpha_n(t)\e^{in\theta}
\hspace{0.7cm}|\alpha_n| \ll 1   \]
Suppose that the direction of the translation is 
along the real axis of the complex plane,
 then the average velocity and the average power expenditure are given by
\begin{eqnarray}
\bar U_{\bar z}&=&{1 \over 2T}\int_0^Tdt \sum_{mn}
\{F_{mn}\alpha_m\dot\alpha_n+F_{m\bar n}
\alpha_m\bar{\dot\alpha}_n+F_{\bar m n}
\bar\alpha_m \dot\alpha_n+F_{\bar m \bar n}\bar\alpha_m
\bar{\dot\alpha}_n\} \nn
\bar{\cal P}&=&{4\pi \mu \over T}\int_0^Tdt \sum_{mn}
P_{mn}\dot \alpha_m \bar{\dot\alpha}_n
\end{eqnarray}
where the field strength tensor are 
\begin{eqnarray}
F_{mn}&=&\{(n+1)\theta_{-n}-(m+1)\theta_{-m}\}\delta_{m+n,-1} \cOa \nn
F_{m\bar n}&=&\{-(n+1)\theta_{-n}+(m+1)\theta_{m}\}\delta_{m-n,1} \cOa\nn
F_{\bar m n}&=&\{-(n+1)\theta_{n}+(m+1)\theta_{-m}\}\delta_{m-n,-1} \cOa\nn
F_{\bar m\bar n}&=&\{-(n+1)\theta_{-n}+(m+1)\theta_{-m}\}
\delta_{m+n,1} \cOa,\nonumber
\end{eqnarray}
and
\[
\theta_n = \left\{ \begin{array}{ll}
1 & (n\geq 0) \\
0 & (n<0) 
 \end{array}   \right. .
\]
The metric tensor is
\begin{equation}
P_{mn}=|n+1|\delta_{m,n} \cOa
\end{equation}
From (\ref{vari}), we obtain 
\begin{equation}
\frac{\delta \eta}{\delta \alpha_m} \propto \sum_n \left( F_{mn}\dot{\alpha}_n
-\lambda P_{mn}\ddot{\alpha}_n \right)= 0  \label{vari2}
\end{equation}
where $\lambda=|-8\pi\mu\eta|$.
If we take the limit of large mode number ($|m|\rightarrow \infty$),
we obtain linear differential equations for $\alpha$  
in the leading order
\begin{equation}
\left(\begin{array}{cc}
2B & B \\
B & -2B \\
\end{array}\right)
\left(\begin{array}{c}
\dot{\bar V}^- \\
\dot{V}^+ \\
\end{array}\right)
=
\lambda
\left(\begin{array}{c}
\ddot{\bar V}^- \\
\ddot{V}^+ \\
\end{array}\right), \label{egncil}
\end{equation}
where the eigenvector is
\[
V^-=\left(\begin{array}{c}
\alpha_{-1} \\ \alpha_{-2} \\ \alpha_{-3} \\ \cdot \\
\cdot \\ \cdot \\ \end{array}\right)
\ , \ \ 
V^+=\left(\begin{array}{c}
\alpha_{1} \\ \alpha_{2} \\ \alpha_{3} \\ \cdot \\
\cdot \\ \cdot \\ \end{array}\right)
\]
and the matrix $B$ is 
\[
B=\left(\begin{array}{cccccccccc}
\cdot & \cdot &&&&&&&&  \\
\cdot & \cdot & \cdot &&&&&&0& \\
& \cdot & \cdot &\cdot &&&&&& \\ 
&& -1 & 0 & 1 &&&&& \\
&&& -1 & 0 & 1 &&&& \\
&&&& -1 & 0 & 1 &&& \\
&&&&& \cdot & \cdot & \cdot && \\
&0&&&&& \cdot & \cdot & \cdot & \\
&&&&&&& \cdot & \cdot & \cdot \\
\end{array}
\right).
\]
When we regard the index $n$ of $\alpha_n$ as the $n$-th lattice point,
the matrix $B$ acts as a difference operator.
The eigenvector of this difference operator is always given by a plane wave.
By solving the eigenvalue equation(\ref{egncil}), we obtain the value 
of the efficiency
\begin{equation}
\lambda= \frac{\sqrt{5}\, T}{\pi} |\sin \beta|,   \label{effcil}
\end{equation}
where $\beta$ is an arbitrary real number.
This gives an upper bound on $\lambda$.\footnote{Though 
the result from Shapere and Wilczek is correct qualitatively,
they made a few mistakes in solving the eigenvalue equation.}
The eigenvector corresponding to the above eigenvalue is 
\[
\alpha_m(t) = \left\{ 
\begin{array}{ll} \e^{i\{ m\beta + m {2\pi \over T}t \} } & m>0 \\
(\pm\sqrt{5} - 2)\e^{i\{m\beta - m {2\pi \over T}t \} }& m<0
\end{array} \right.
\]
Though our calculation has been limited to large mode number,
it can be shown that including the finite mode number does not bring about 
more efficient stroke of \mo \ in two dimensional fluid
(See Appendix in \cite{wil2}).
In fact, a real ciliate is covered with thousands of short cilia, and 
this means that the mode number of the surface of the ciliate is 
extremely high.

%%%%%%%%%%%%%%%%%%%%%%%%%%%%%%%%%%%%%%%%%%%%%%%%%%%%%%%%%%%%%%%%%%%5
\subsection{Flagellated motion}

We regard the shape of a flagellate as an infinitesimal deformation of 
a segment of a line.
A convenient choice is a parametrization of the shape such that
\begin{equation}
S(t, \theta) = 2 \{ \cos \theta + i \sin \theta \alpha(t,\theta) \},
\end{equation}
where 
\[
\alpha(t,\theta)=\sum_{n=1}^\infty \alpha_n(t) \sin n\theta,
\hspace{0.7cm}|\alpha(t,\theta)| \ll 1, \hspace{0.7cm}
-\pi \leq \theta \leq \pi
\]
and suppose that $\alpha_n$ is a real number so that the length of flagellate 
can be locally preserved at ${\cal O}(\alpha)$.
Fortunately, this requirement eliminates the possibility of rotational motion.
In this case, by making use of the result from Ref.\cite{ma},
the field strength tensor and the 
metric tensor evaluated at $S_0=2\cos \theta$ are 
\begin{eqnarray}
F_{mn} & = & (2n+1) \delta_{n-m,-1} -(2m+1)\delta_{n-m,1}  \cOa
\nonumber \\
P_{mn} & = & 2n\delta_{n,m} -(n+1)\delta_{n-m,-2} -(n-1)\delta_{n-m,2}
 \cOa.
\label{metfla}
\end{eqnarray}
By calculating (\ref{vari}) again, in the limit of large mode number,
we obtain an eigenvalue equation of the following form:
\begin{equation}
\left(\begin{array}{cc}
0 & - B \\ B & 0 \\
\end{array}\right)
\left(
\begin{array}{c}
\dot V_{{\rm even}} \\
\dot V_{{\rm odd}} \\
\end{array}\right)
=
\lambda
\left(\begin{array}{cc}
-{1 \over 2}B^2 & 0 \\ 0 & -{1 \over 2}B^2 \\
\end{array}\right)
\left(
\begin{array}{c}
\ddot V_{{\rm even}} \\
\ddot V_{{\rm odd}} \\
\end{array}\right),      \label{egnfla}
\end{equation}
where
\[
V_{{\rm even}}=\left(\begin{array}{c}
\alpha_2 \\ \alpha_4 \\ \alpha_6 \\ \cdot \\
\cdot \\ \cdot \\ \end{array}\right)
\ , \ \ 
V_{{\rm odd}}=\left(\begin{array}{c}
\alpha_{1} \\ \alpha_{3} \\ \alpha_{5} \\ \cdot \\
\cdot \\ \cdot \\ \end{array}\right)
\]
By solving (\ref{egnfla}), we obtain the eigenvectors
\begin{equation}
\alpha_n(t) = \left\{ 
\begin{array}{ll}
\sin\,m\beta'\cos{2\pi t \over T} & m:{\rm even}\\
 \pm \cos\,m\beta'\sin{2\pi t \over T} &  m:{\rm odd} 
\end{array} \right.
\end{equation}
with the eigenvalues 
\begin{equation}
\lambda={T \over 2\pi|\sin \beta'|},   \label{efffla}
\end{equation}
where $\beta'$ is an arbitrary real number.
Notice that this reslut gives no upper bound on $\lambda$.
This unboundedness is caused by the fact that the square of the difference 
operator $B^2$ appears in the right-hand side of (\ref{egnfla}).
It indicates naively that the efficiency of the swimming of flagellate 
can be much better than that of ciliate 
at Reynolds number zero.
In other words, flagellates adjust themselves to the sticky world ($Re=0$)
more efficiently than ciliates.
This might be the reason why in general, the size of flagellate 
is smaller than that of ciliate.

%%%%%%%%%%%%%%%%%%%%%%%%%%%%%%%%%%%%%%%%%%%%%%%%%%%%%%%%%%%%%%%%%%
\section{Discussion}

The reason why there is no upper bound on the efficiency of flagellate is 
intimately tied up with the structure of the shape space.
From eq.(\ref{metfla}), the matrix $P_{mn}$ is proportional to a non-local 
operator $B^2$, which has zero mode, in the limit of large mode number.
This means that the shape space becomes singular at the basic shape of 
flagellate
\[
S_0(\theta)=2\cos\theta,
\]
where the dimensions of the shape space are reduced if we regard $P_{mn}$ as 
a metric tensor on the space.
It turns out that the basic shape of the flagellate is a boundary 
of the shape space.
Assuming that the time dependence of the eigenvector is, 
as we have done in previous sections, given by 
\[
\alpha_m(t)\propto \e^{i{\rm sign}(m){2\pi \over T}t}
\hspace{1cm}{\rm for\  ciliate}, \]
\[
\alpha_n(t)\propto \cos{2\pi t \over T}, \ \sin{2\pi t \over T}  
\hspace{1cm}{\rm for\  flagellate},
\]
the average power expenditure is expressed as 
\begin{eqnarray}
\bar{\po} &=&{1 \over T}\int_0^T \po dt \nn
&\propto&{1 \over T^2}
\left( \int_0^T \sqrt{\po}dt \right)^2 \nonumber
\end{eqnarray}
One cycle of the deformation of the shape corresponds to a closed path 
in the \ss.
Since $\sqrt{\po}dt$ is the infinitesimal line element in the shape space,
$\bar{\po}$ turns out to be proportional to the square of perimeter 
of the closed path in the \ss.
On the other hand, the average velocity
\[
\bar U^i ={1 \over T}\int_0^T
F^{T^i}_{\omega_m\omega_n}\alpha_m\dot\alpha_n dt 
\]
is proportional to the area in the shape space enclosed by the closed path.
Thus, roughly speaking, our definition of the \ef\ of \mo\ (\ref{defeff})
is provided by the ratio of the area in the \ss\ enclosed by the circular path 
to its perimeter squared.
As the value of the \ef\ is larger, the perimeter squared becomes 
much smaller than its area. Therefore, 
the shape space is sharpened like a needle around 
the basic shape of flagellate.
Our study would be the first step towards understanding 
what the global structure of \ss \ is like.

%%%%%%%%%%%%%%%%%%%%%%%%%%%%%%%%%%%%%%%%%%%%%%%%%%%%%%%55
\section{Summary}

The issue of the efficiency of the swimming of \mo \ has provided us 
various interesting results. 
The efficiency of the swimming of flagellate was of particular interest,
because the value of the efficiency has no upper bound.
While the metric tensor of the \ss\ seems to be locally flat 
around the basic shape of ciliate,
a singularity is found at the basic shape of 
flagellate where the power expenditure becomes zero.
This fascinating structure of the \ss \ reflects the 
unboundedness of the efficiency of flagellate.

By the same token, our treatment in this paper can be generalized to the 
case of \mo \ in three dimensional fluid.
Shapere and Wilczek also found maximally swimming strokes for nearly 
spherical organism whose deformation is azimuthally symmetric. 
The organism corresponds to the ciliate in three dimensional 
fluid\cite{wil2}.
Remarkably, the result for the sphere and cylinder in three dimensional fluid 
were identical in the limit of large mode number. 
The latter corresponds to ciliate in two dimensional fluid.
Although the treatment of swimming motion of flagellate and bacterium 
with bacterial flagella  
in three dimensional fluid is likely to be more complicated, 
it would be interesting to explore the problem on 
the hierarchy of the size and the structure 
of the \ss.

%%%%%%%%%%%%%%%%%%%%%%%%%%%%%%%%%%%%%%%%%%%%%%%%%%%%%%%%%%%%%%%%%%
\section*{Acknowledgments}

I am grateful to S. Nojiri for valuable discussions and 
his collaboration at the early stage of this work.
I would also like to thank A. Sugamoto for useful discussions.

%%%%%%%%%%%%%%%%%%%%%%%%%%%%%%%%%%%%%%%%%%%%%%%%%%%%%%%%%%%%%%%%%%

\end{document}